%% file: paper.tex
\pdfoutput=1

\newif\ificdcs
\icdcsfalse

\ificdcs
\documentclass[10pt,conference,compsocconf,letterpaper]{IEEEtran}
\else
\documentclass[10pt,conference,compsocconf,letterpaper]{IEEEtran}
\fi

\usepackage{graphics}
\usepackage{epsfig}
\usepackage{subfigure}
\usepackage{color}
\usepackage{url}
\usepackage{amsfonts}
\usepackage{multirow}
\usepackage{xspace}
\usepackage{amsmath} 
\usepackage[lined,linesnumbered,vlined,figure]{algorithm2e}

\usepackage[bookmarks=false, colorlinks=true, citecolor=blue, colorlinks=blue, linkcolor=blue, urlcolor=blue]{hyperref}

\usepackage[letterpaper, textwidth=7in, left=0.73in, right=0.73in, top=0.8in, bottom=0.8in]{geometry}

\begin{document}

\title{Tiresias: Online Anomaly Detection for Hierarchical Operational Network Data
}

\author{\IEEEauthorblockN{Chi-Yao Hong, Matthew Caesar}
\IEEEauthorblockA{University of Illinois at Urbana-Champaign\\
\{cyhong$|$caesar\}@illinois.edu}
\and
\IEEEauthorblockN{Nick Duffield, Jia Wang}
\IEEEauthorblockA{AT\&T Labs - Research\\
\{duffield$|$jiawang\}@research.att.com}
}


\maketitle

\newcommand{\mattc}[1]{{\color{red}{#1}}}
\newcommand{\matt}[1]{{\color{blue}{#1}}}

\newcommand{\paragraphb}[1]{\vspace{0.03in}\noindent{\bf #1} }
\newcommand{\paragraphe}[1]{\vspace{0.03in}\noindent{\em #1} }
\newcommand{\paragraphbe}[1]{\vspace{0.03in}\noindent{\bf \em #1} }
\newcommand{\comment}{\textcolor{black}}
\newcommand{\mcomment}{\textcolor{red}}
\newcommand{\ecomment}{\textcolor{black}}
\newcommand{\newcomment}{\textcolor{black}}
\newcommand{\cy}{\textcolor{magenta}}
\newcommand{\SYNDEG}{\mbox{\textsf{SYN\_DEG}}}
\newcommand{\us}{\,$\mu$s\xspace}
\newcommand{\systh}{\verb"123"\xspace}

\renewcommand{\textsf}{}
\newcommand{\sys}{\textsf{Tiresias}\xspace}
\newcommand{\sysnp}{\textsf{Tiresias}\xspace}
\newcommand{\bfa}{\textsf{BA1}\xspace}
\newcommand{\bfb}{\textsf{STA}\xspace}
\newcommand{\alg}{\textsf{ADA}\xspace}
\newcommand{\opg}{\textsf{OP}\xspace}
\newcommand{\negsp}{\!}

\newcommand{\tracea}{\textsf{CCD}\xspace}
\newcommand{\tracean}{\textsf{CCD}\xspace}
\newcommand{\traceb}{\textsf{CCD2}\xspace}
\newcommand{\tracebn}{\textsf{CCD2}\xspace}
\newcommand{\tracec}{\textsf{SCD}\xspace}
\newcommand{\tracecn}{\textsf{SCD}\xspace}

\newcommand{\heu}{\textit{Dummy\_Layer}$(h)$\xspace}

\newcommand{\SYN}{\mbox{\textsf{SYN}}}
\newcommand{\SYNRSP}{\mbox{\textsf{SYN-RSP}}}

\newtheorem{lem}{Lemma}
\newtheorem{col}{Corollary}
\newtheorem{defi}{Definition}

\input{abstract}
\input{intro}
\input{measurement}

\input{problem}

\input{algorithm}

\input{timeseries}

\input{evaluation}
\input{related}

\input{conclusion}


\bibliographystyle{IEEEtran}
\bibliography{paper}

\end{document}

%% file: abstract.tex
\begin{abstract}

\emph{Operational network data}, management data such as customer care call logs and
equipment system logs, is a very important source of information for network
operators to detect problems in their networks.  Unfortunately, there is lack of
efficient tools to {\em automatically} track and detect anomalous events on
operational data, causing ISP operators to rely on manual inspection of this
data.  While anomaly detection has been widely studied in the context of
network data, operational data presents several new challenges, including the
volatility and sparseness of data, and the need to perform fast detection
(complicating application of schemes that require offline processing or
large/stable data sets to converge).

To address these challenges, we propose \sys, an automated approach to locating
anomalous events on hierarchical operational data.  \sys leverages the
hierarchical structure of operational data to identify high-impact aggregates
(e.g., locations in the network, failure modes) likely to be associated with
anomalous events.  To accommodate different kinds of operational network data,
\sys consists of an online detection algorithm with low time and space
complexity, while preserving high detection accuracy.  We present results from
two case studies using operational data collected at a large commercial IP network operated by a Tier-1 ISP:
customer care call logs and set-top box crash logs.  By comparing with a reference set
verified by the ISP's operational group, we validate that \sys can achieve
$>$$94\%$ accuracy in locating anomalies. \sys also discovered several previously
unknown anomalies in the ISP's customer care cases, demonstrating its
effectiveness.


\end{abstract}

%% file: intro.tex
\section{Introduction}

Network monitoring is becoming an increasingly critical requirement for 
protecting large-scale network infrastructure. Criminals and other Internet
miscreants make use of increasingly advanced strategies to propagate spam,
viruses, worms, and exercise DoS attacks against their targets~\cite{porras:leet09, Qian:oakland:10, stover:login07}. 
However, the growing size and functionality of modern network deployments
have led to an increasingly complex array of failure modes and anomalies
that are hard to detect with traditional approaches~\cite{Choffnes10, Mahimkar08}.

Detection of statistically anomalous behavior is an important line of defense for protecting large-scale network infrastructure. 
Such techniques work by analyzing {\em network traffic} (such as traffic volume, communication outdegree, and rate of routing updates), characterizing ``typical'' behavior, and
extracting and reporting statistically rare events as alarms~\cite{Brauckhoff:imc09, autofocus, lakhina:sigcomm04, lakhina:sigcomm05, Zhang:mnd05}.  

%

While these traditional forms of data are useful, the landscape of ISP
management has become much broader. Commercial ISPs collect an increasingly
rich variety of {\em operational data} including network trouble tickets,
customer care cases, equipment system logs, and repair logs.  
%
%
Several properties of operational data make them highly useful for network
monitoring.
Operational data is often annotated with semantics.  Customer care
representatives annotate call logs with detailed characteristics on the end
user's experience of the outage, as well as information on the time and
location of the event.  Repair logs contain information about the underlying
root cause of the event, and how the problem was repaired.  
This provides more detailed information on the meaning and the type of event that
occurred from a smaller amount of data. More importantly, operational data is often
presented in an explicitly-described hierarchical format.  
While the underlying signal may be composed across multiple dimensions, correlations across those
dimensions are explicitly described in the provided hierarchy, allowing us to
detect anomalies at higher levels of the tree when information at the leaves is
too sparse.

%
%

While operational data can serve as rich sources of information
to help ISPs manage network performance, there is still a lack of an effective
mechanism to assist ISP operation teams in using these data sources to identify network anomalies.  
For example, currently a Tier-$1$ ISP operation group \emph{manually} inspects
$>$$300$,$000$ customer care calls in a working day to deduce network
incidents or performance degradations.  Unfortunately, the sheer volume of data
makes it difficult to manually form a ``big picture'' understanding of the
data, making it very challenging to locate root causes of performance problems accurately.





To deal with this, it is desirable to {\em automatically} detect anomalies
in operational data.
Unfortunately, several unique features of operational data complicate application
of previous work in this space:\\
\textbf{Operational data is complex:} Operational data is {\em
sparse} in the time domain, with a low rate of underlying events; it is
composed of data spread over a \emph{large hierarchical} space, reducing the magnitude of
the signal.  Operational data is also {\em volatile}, being sourced from a data
arrival rate that changes rapidly over time, complicating the ability to detect
statistical variations.  Unfortunately, the fact that user calls arrive on the
order of minutes for a particular network incident (as opposed to sub-millisecond reporting in traditional
network traffic monitoring systems) means that the ISP may need to make a
decision on when to react after a few calls.\\
\textbf{Operational data represents urgent concerns:} Operational
data contains mission-critical failure information, often sourced {\em after}
the problem has reached the customer.  Operational data such as call logs
reflects problems crucial enough to motivate the customer to contact the call
center.  Moreover, call logs that affect multiple customers typically reflect
large-scale outages that need immediate attention.  Unfortunately, the richness
of operational data incurs significant processing requirements,
complicating use of traditional anomaly detection approaches (which rely on
offline processing).

To address these challenges, we propose \sys \footnote{The name comes from the intuition 
that our system detects network anomalies without needing to understand the context of operational logs.}, an automated approach to locating
anomalous events (e.g., network outages or intermittent connection drops) on operational data.
To leverage the hierarchical nature of
operational data, we build upon previous work on {\em hierarchical heavy
hitter}-based anomaly detection~\cite{autofocus,Zhang:imc04}. 
In a large hierarchy space, heavy hitters are nodes with higher aggregated data counts, which 
are suspicious locations where anomalies (significant spikes) could happen.
However, these works relied on offline processing for anomaly detection, which
does not scale to operational data (refer to \S\ref{sec:related} for further
details).  We extend these works to perform anomaly detection in an {\em
online} and {\em adaptive} fashion.  To quickly detect anomalies without
incurring high processing requirements, we propose novel algorithms to
maintain a time series characterization of data at each heavy hitter that is
updated in both time and hierarchical domain dynamically without requiring
storage of older data.  To avoid redundant reports of the same anomaly, our
approach also automatically discounts data from descendants that are
themselves heavy hitters.

\ificdcs
\else
We first characterize two large-scale operational datasets (customer care calls and set-top boxes crash logs; \S\ref{sec:measurement}) via measurement.
Then we formally define our problem (\S\ref{sec:problem}).
We present a high-level overview of our design and its limitations (\S\ref{sec:overview}).
To achieve online detection, we propose novel algorithms to 
update the time series in both time and hierarchical domain dynamically (\S\ref{sec:alg}).
We adopt time series analysis to derive the data seasonality and perform time series based anomaly detection (\S\ref{sec:tsa}).
We evaluate the efficiency and effectiveness of \sys using large-scale operational datasets (\S\ref{sec:evaluation}). 
Our findings suggest that \sys can scale to these real datasets with sufficiently low computational resources (\S\ref{sec:eval:alg}).
We compare the set of anomalies detected by \sys on customer call cases with an existing approach 
used by a major US broadband provider~(\S\ref{sec:eval:dem}). \sys discovers $>$$94\%$ of the 
anomalies found by the reference method as well as many previously unknown anomalies hidden in the lower levels. 
We present related work (\S\ref{sec:related}) before we conclude (\S\ref{sec:conclusion}).
\fi

%% file: measurement.tex
\section{Characterizing Operational Data}
\label{sec:measurement}

It is arguably impossible to accurately detect anomalous events without understanding the underlying data.
We first present characterization results of two operational datasets (\S\ref{sec:measurement:dataset}).
Then we conduct a measurement analysis to study the properties of the operational datasets (\S\ref{sec:measurement:property}).


\begin{figure*}[t]
\centering
\subfigure[]{\label{fig:prop:rate:uverse:td}{\includegraphics[width=2.25in]{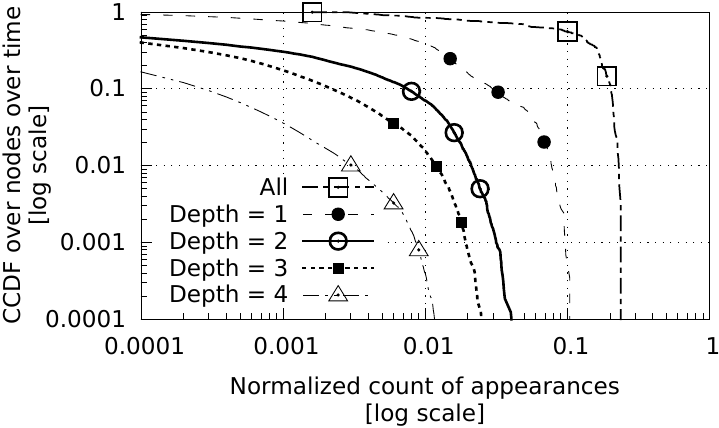}}}\hspace{.08in}
\subfigure[]{\label{fig:prop:rate:uverse:nw}{\includegraphics[width=2.25in]{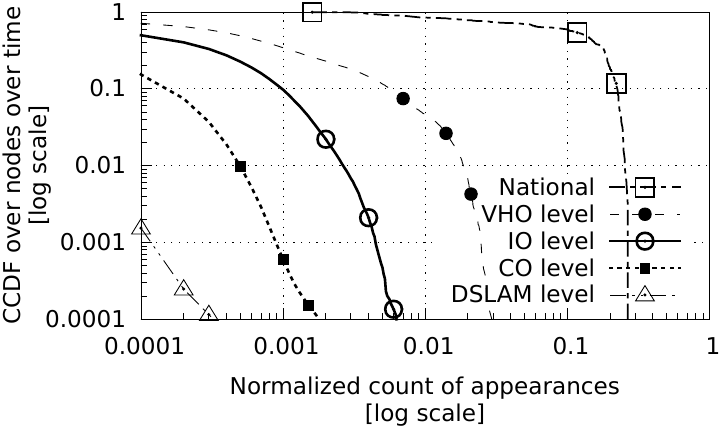}}}\hspace{.08in}
\subfigure[]{\label{fig:prop:rate:stb}{\includegraphics[width=2.25in]{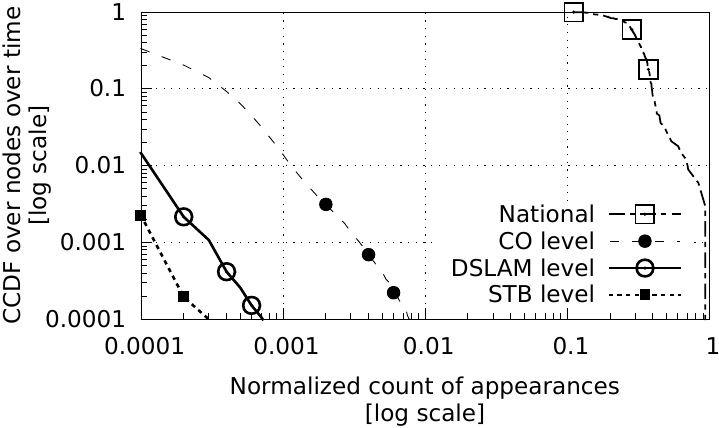}}}
\vspace{-5pt}
\caption{
For each level in the hierarchy, the complementary cumulative distribution 
function of the normalized count of cases across nodes and time units. 
(a) trouble issues of \tracea, (b) network 
locations of \tracea, and (c) network locations of \tracec.
Note that both x- and y-axes are log scale. 
}
\vspace{-10pt}
\label{fig:prop:rate}
\end{figure*}

\subsection{Operational Datasets}
\label{sec:measurement:dataset}

The data used in this study was obtained from a major broadband
provider of bundled Internet, TV and VoIP services in the US. 
We present the results based on the datasets\footnote{all customer
identifying information was hashed to a unique customer index to protect privacy.} collected from May to September $2010$.
Two distinct operational datasets were used,
which we now describe.


\paragraphb{Customer Care Call Dataset (\tracea):} The study used data
derived from calls to the broadband service customer care center.
A record is generated for each such call. Based on the conversation with the customer, the care 
representative assigns a category to the call, choosing from
amongst a set of \emph{predetermined hierarchical} classes.
We focus on analyzing customer calls that are related to performance troubles
(service disruptions/degradation, intermittent connectivity,
etc.) and discard records unrelated to performance
issues (e.g., calls about provisioning, billing, accounting).
The resulting operator-assigned category constitutes a \emph{trouble description}, 
which is drawn from a hierarchical tree with a height of $5$ levels\footnote{These categories indicate the broad area of the trouble: IPTV, home
wireless network, Internet, email, digital home phone, and a more
specific detail within the category.
For example, {[\texttt{Trouble} \texttt{Management}]}
~$\Rightarrow$~[{\texttt{TV}}]
~$\Rightarrow$~[{\texttt{TV} \texttt{No} \texttt{Service}}]
~$\Rightarrow$~[{\texttt{No} \texttt{Pic} \texttt{No} \texttt{Sound}}]
~$\Rightarrow$~[{\texttt{Dispatched} \texttt{to} \texttt{Premise}}].}.
We summarize the distribution of the number of customer tickets
for the first-level categories (Table~\ref{tbl:td_issues}).
Each record also specifies information concerning the \emph{network path} to 
the customer premises, which is drawn from a hierarchical
tree comprising $5$ levels\footnote{The top (root) level is SHO (Super Head-end
Office), which is the main source of national IPTV content. Under
SHO, there are multiple VHOs (Video Head-end Office) in the second
level, each of which is responsible for the network service in a
region. A VHO serves a number of IOs (intermediate offices), each of
which serves a metropolitan area network.  An IO serves multiple
routers/switches called COs (central offices), and each CO serves
multiple switches DSLAMs (digital subscriber line access multiplexers)
before reaching a customer residence.}.

\begin{table}[t]
\caption{\tracean customer calls}
\vspace{-9pt}
\centering
\begin{tabular}{|l|c|}\hline
\textbf{Ticket Types} & \textbf{Percentage~($\%$)} \\ \hline\hline
TV                              & $ 39.59  $ \\ \hline
All Products                    & $ 26.71  $ \\ \hline
Internet                        & $ 10.04  $ \\ \hline
Wireless                        & $  9.26  $ \\ \hline
Phone                           & $  8.46  $ \\ \hline
Email                           & $  3.59  $ \\ \hline
Remote Control                  & $  2.35  $ \\ \hline
\end{tabular}
\label{tbl:td_issues}
\end{table}

\begin{table}[t]
\caption{Hierarchy properties}
\vspace{-9pt}
\centering
\setlength{\tabcolsep}{5.3pt}
\begin{tabular}{|c|c|c|c|c|c|c|c|}\hline
\multirow{2}{*}{\textbf{Data}} & \multirow{2}{*}{\textbf{Type}} & \multirow{2}{*}{\textbf{Depth}} & \multicolumn{4}{c|}{\textbf{Typical degree at $k$th level}} \\ \cline{4-7}
 & & & $k=1$ & $k=2$ & $k=3$ & $k=4$ \\ \hline\hline
\multirow{2}{*}{~\tracean} & Trouble descr.    & $5$ & $9$ & $6$ & $3$ & $5$  \\ \cline{2-7}
                         & Network path & $5$ & $61$ & $5$ & $6$ &$24$  \\ \hline
\tracecn                  & Network path & $4$ & $2$,$000$ & $30$ & $6$ & N/A \\ \hline
\end{tabular}
\vspace{-13pt}
\label{tbl:nodes}
\end{table}


\begin{figure*}[t]
\centering
\subfigure[]{\label{fig:prop:ts:uverse}{\includegraphics[width=3in]{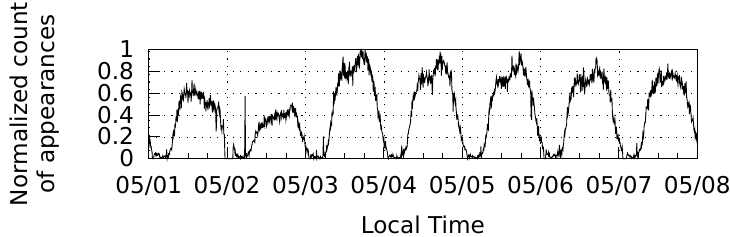}}}
\hspace{18pt}
\subfigure[]{ \label{fig:prop:ts:stb}{\includegraphics[width=3in]{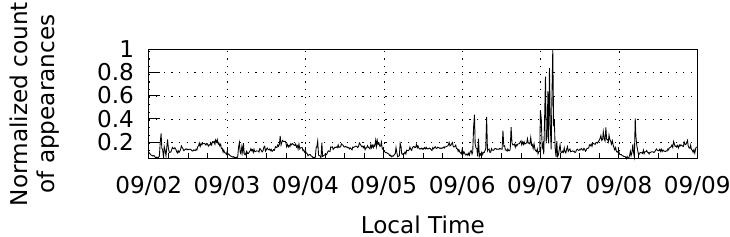}}}
\vspace{-8pt}
\caption{Two representative time series of the count of appearances in a $15$-minute precision for the datasets (a) \tracea and (b) \tracec. The data count is normalized, by dividing it by the maximal count in the time series. (a) and (b) starts from Saturday and Thursday, respectively.}
\vspace{-15pt}
\label{fig:prop:ts}
\end{figure*}



\paragraphb{STB Crash Dataset (\tracec):}
At the customer premises, a Residential Gateway serves as a modem and
connects to one or more Set-Top Boxes (STBs) that serve TVs and enable
user control of the service. Similar to other types of computer
devices, an STB may crash, then subsequently reboot. This study used
logs of STB crash events from the same broadband network, that are
generated by the STB. By combining the STB identifier reported in the log, with
network configuration information, network path information similar to
that described for \tracea can be associated with each crash event.
This information is represented in a hierarchical tree of $4$ levels. 
The broad properties such as node degrees in different levels are summarized (Table~\ref{tbl:nodes}).


\subsection{Characteristics of Operational Data}
\label{sec:measurement:property}

\paragraphb{Sparsity:} We measure the case arrival rate for difference aggregates in the hierarchies (Fig.~\ref{fig:prop:rate}).
The number of cases per node, especially for nodes in the lower levels, is very sparse over time in both data sets. 
For example, we observe that in about $93\%$ ($70\%$) of the time we have no cases for nodes in the CO level in \tracea (\tracec). 
Anomaly detection over sparse data is usually inaccurate as it is hard to differentiate the anomalous and 
normal case~\cite{botgraph}. However, larger localized bursts of calls relating to certain nodes do
occur; these are obvious candidates for anomalies.

\paragraphb{Volatility:} In both datasets, we observe the number of cases varies significantly over time. 
For example, consider the root node of trouble issues in \tracea (``All'' in Fig.~\ref{fig:prop:rate:uverse:td}), 
we observe that the case number at the $90^{\mathrm{th}}$ percentile 
of the appearance count distribution is $\approx\negsp35$ times higher than that at the $10^{\mathrm{th}}$ percentile.
We observe that the sibling nodes (nodes that share a common parent) in the hierarchy 
could have very different case arrival rates. 
The above observations imply that the set of heavy hitters could change significantly over time. 
Therefore, observing any fixed subset of nodes in the hierarchy could easily miss significant anomalies.


\paragraphb{Seasonality:} To observe the variation on arrival case counts, 
we measured the time series of the normalized count of appearances in $15$-minute time units (Fig.~\ref{fig:prop:ts}). 
Overall, a diurnal pattern is clearly visible in both the datasets. 
This pattern is especially strong in \tracea.
In particular, the daily peaks are usually around $4$ PM, while the daily minimum is at around $4$ AM.
Moreover, the weekly pattern is observed in \tracea where we have fewer cases in first two days (May $1-2$, 2010), 
which are Saturday and Sunday (Fig.~\ref{fig:prop:ts}). 
Regardless of the differences observed in the two datasets, one common property is the presence of large spikes. 
Some spikes last only a short period (e.g., $<$$30$ minutes at $5$ AM, May $2$nd in Fig.~\ref{fig:prop:ts:uverse}), while 
some spikes last longer (e.g., $>$$5$ hours at off-peak time, Sept $7$th in Fig.~\ref{fig:prop:ts:stb}).
These spikes are by no means unique over time -- we observed a similar pattern in other time periods. 


%% file: problem.tex
\begin{figure*}[t]
\centering
\includegraphics[width=6in]{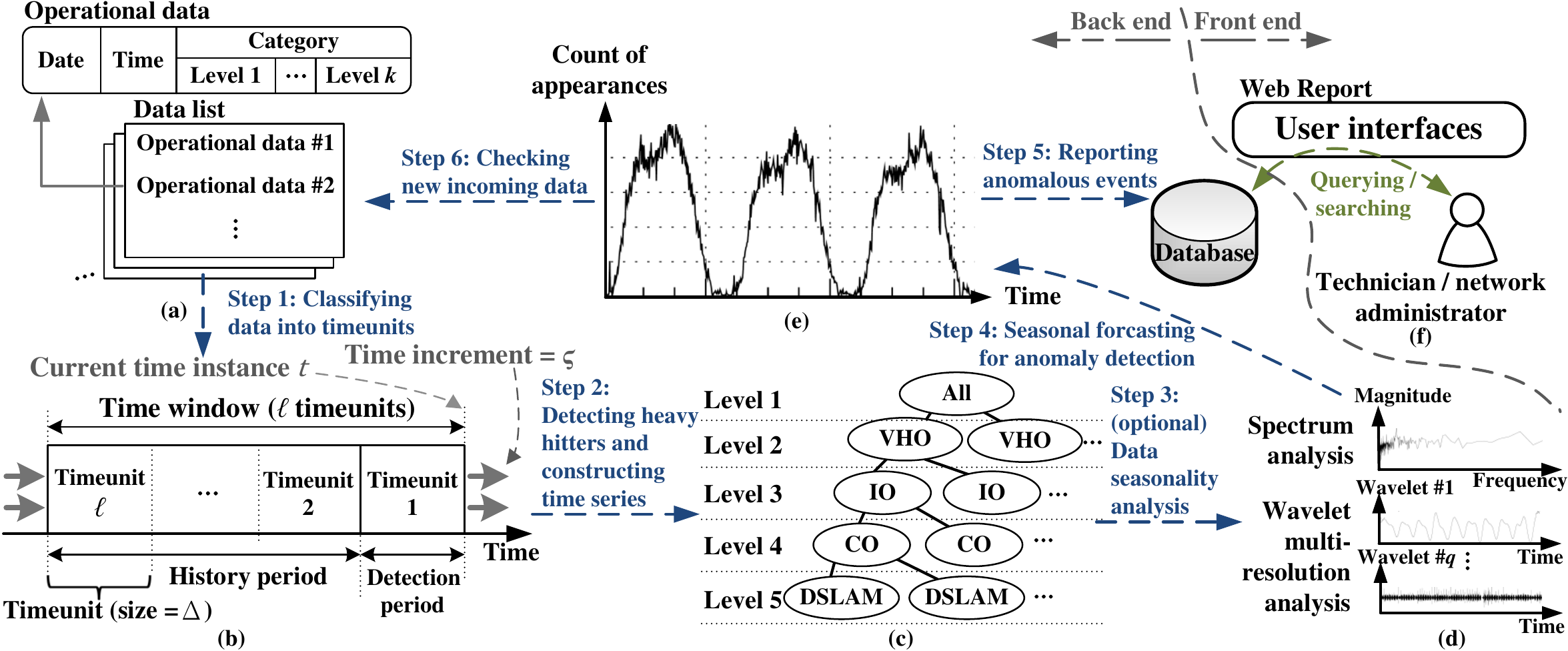}
\vspace{-10pt}
\caption{
System diagram: (a) abstraction of streaming operational data, (b) a window-based structure to classify input data into time units, 
(c) heavy hitter detection and time series construction in a hierarchical domain, (d) data seasonality analysis, (e) seasonal forecasting 
model for anomaly detection, and (f) a user interface enables an administrator to query for anomalous events.
For details, refer to \S\ref{sec:overview}.
}
\vspace{-15pt}
\label{fig:sys}
\end{figure*}

\section{Problem Statement}
\label{sec:problem}


In this section, we formally define the online anomaly detection problem. 
In particular, we first describe the operational data format, and then we define \emph{hierarchical heavy hitter} which 
allows us to identify the high-impact aggregates. To detect if the event related to these heavy hitters is anomalous, we 
define \emph{time series based anomaly detection}. 


An input stream $\textsf{S} = s_0, s_1, \dots$ is a sequence of operational network data $s_i$.
Each data $s_i = (k_i, t_i)$ consists of two components: \emph{category} $k_i$ provides classification 
information about the data; \emph{time} $t_i$ denotes the recorded date and time of $s_i$.
The data can be classified into ``timeunits'' based on its time information $t_i$. 
The data category is drawn from a hierarchical domain (i.e., tree) where each node $n$ in the tree is associated 
with a \emph{weight} value $A_n[k, t]$ for timeunit $t$.
For leaf nodes, the weight $A_n[k, t]$ is defined by the number of data items 
in $\textsf{S}$ whose category is $k$ and occur in timeunit $t$.
For non-leaf nodes (i.e., interior nodes), the weight $A_n[k, t]$ is the summation 
of the weights of its children. We defined the \emph{hierarchical heavy hitter (HHH)} as follows:
\begin{defi}\label{def:hhh}
\normalfont (HHH; Hierarchical Heavy Hitter). Consider a hierarchical tree where each node $n$ is weighted by $A_n[k,t]$ for timeunit $t$. 
Then the set of HHH for timeunit $t$ can be defined by $\textsf{HHH}[\theta, t] = \{n | A_n[k, t] \ge \theta \}$ for a threshold $\theta$.
\end{defi}
These heavy hitters (i.e., high-volume aggregates) are suspicious locations where anomalies (i.e., significant spikes) could 
happen\footnote{Intuitively, if an aggregate has very little data in a timeunit, it is unlikely to see a huge spike there. 
We expect to capture most anomalies by setting a sufficient small heavy hitter threshold.}.

In Definition~\ref{def:hhh}, the count of appearances of any leaf node will contribute to \emph{all} its ancestors.
Therefore, if we perform anomaly detection without taking the hierarchy into account, an 
anomaly in a leaf node may be reported multiple times (at each of its ancestors).
To address this, one could reduce the redundancy by removing a heavy hitter node $n$ if its weight can be inferred 
from that of heavy hitter node $n_d$ that is a descendant of $n$. 
To arrive at this more ``compact'' representation, we consider a more strict definition of HHHs to find nodes 
whose weight is no less than $\theta$, \emph{after discounting the weights from descendants that are themselves HHHs}.

\begin{defi}\label{def:shhh}
\normalfont (SHHH; Succinct Hierarchical Heavy Hitter). Consider a hierarchical tree where each node $n$ is associated with a \emph{modified weight} value $W_n[k, t]$ for timeunit $t$. 
For leaf nodes, the modified weight $W_n[k, t]$ is equal to the original weight $A_n[k, t]$. 
For non-leaf (internal) nodes, the modified weight $W_n[k, t]$ is the summation of weights of its children that are themselves \emph{not} a heavy hitter;
In other words, given the set of SHHHs, denoted by $\textsf{SHHH}[\theta, t]$, the modified weight for non-leaf nodes can be derived by
\begin{equation*}
W_n[k, t] = \sum_{m} W_{m}[k, t] : m \in child(n) \cap m \notin \textsf{SHHH}[\theta, t]
\end{equation*}
and the succinct hierarchical heavy hitter set is defined by $\textsf{SHHH}[\theta, t] = \{n | W_n[k, t] \ge \theta \}$.
\end{defi}

\ificdcs
\else
Notice that both the set of SHHHs and the modified weights are unique, i.e., there
can be only one heavy hitter set satisfying the above recursive definition\footnote{Suppose that more than one solution exists.
Then there must exist one node $n$ who is in one set $\textsf{SHHH}^{(1)}$ but not in another set $\textsf{SHHH}^{(2)}$.
Then we know $W_n^{(1)} \neq W_n^{(2)}$. By definition, there must exist a node $m \in child(n)$ where $W_m^{(1)} \neq W_m^{(2)}$.
Repeating the above procedures, we end up having a \emph{leaf} node $p$ where $W_p^{(1)} \neq W_p^{(2)}$, which violates the
definition of the modified weight for leaf nodes. 
}.
An intuitive explanation is that the question of whether each node joins the set only depends on the weight of its descendants.
It is not hard to see that a bottom-up traversal on the hierarchy would give the correct set.
We will use this definition throughout this paper (for brevity, we will write ``heavy hitter'' instead of ``succinct hierarchical heavy hitter'' hereafter).
\fi

To detect anomalous events, we will construct forecasting models on heavy hitter nodes to detect anomalous events.
As heavy hitter nodes could change over time instances, we need to construct the time series 
only for nodes which are heavy hitters in the \emph{most recent} timeunit.

\begin{defi}\label{def:cts}
\normalfont (Time Series). For each timeunit $t$, consider a hierarchical tree where each node $n$ is weighted by $A_n[k, t]$.
Given the heavy hitter set in the \emph{last} timeunit ($t=1$), i.e., $\textsf{SHHH}[\theta, 1]$, the time series of 
any heavy hitter node $n \in \textsf{SHHH}[\theta, 1]$ is denoted by $T[n, t] = A_n[k, t] - \sum_{m} A_{m}[k, t] : m \in child(n) \cap m \in \textsf{SHHH}[\theta, 1]$, for $1 \le t \le \ell$, where $\ell$ is the length of time series.
\end{defi}


\begin{defi}\label{def:ad}
\normalfont (Anomaly). Given a time series $T[n, t]$ of any node $n$, $1 \le t \le \ell$, the
forecast traffic of the latest time period $T[n, 1]$ is given by $F[n, 1]$, derived by some forecasting model.
Then there is an anomalous event happening at node $n$ in the latest time period iff $T[n, 1] / F[n, 1] > RT$ and $T[n, 1] - F[n, 1] > DT$, 
where $RT$ and $DT$ are sensitivity thresholds, which are selected by sensitivity test (\S\ref{sec:evaluation}).
We consider both absolute and relative differences to minimize false detections at peak and dip time.

\end{defi}

\section{TIRESIAS System Overview}\label{sec:overview}


We present our system diagram in Fig.~\ref{fig:sys}. For each successive time period 
we receive a data list consisting of a set of newly arrived operational data (Fig.~\ref{fig:sys}(a)). \\
\textbf{Step $1$:} We maintain a sliding time window to group the input streaming data into ``timeunits'' based 
on its time information $t_i$ (Fig.~\ref{fig:sys}(b)). In particular, a time window consists of $\ell$ units, each of which is of size $\Delta$. 
When new data lists arrived, we simply \emph{shift} the time instance / window (by a configurable offset $\varsigma$) to keep track of newly available data.
For each time instance / window, we focus on detecting anomalous events 
in the \emph{last} timeunit (the detection period in Fig.~\ref{fig:sys}(b)), 
while the data in other timeunits is used as the history for forecasting (the history period in Fig.~\ref{fig:sys}(b)).\\
\textbf{Step $2$:} Based on the category information $k_i$ on the data, we construct a classification tree (Fig.~\ref{fig:sys}(c)), 
where each category $k_i$ can be bijectively mapping to a \emph{leaf node} in the tree. As nodes associated with higher aggregated 
data counts in the last timeunit are suspicious places where anomaly could happen, 
we first detect the heavy hitter set in the classification tree~(Fig.~\ref{fig:sys}(c); Definition \ref{def:shhh}).
Then we construct time series for these heavy hitters (Definition \ref{def:cts}).
We will present efficient algorithms to achieve these goals~(\S\ref{sec:alg}).\\
\textbf{Step $3$:} We perform data seasonality analysis (Fig.~\ref{fig:sys}(d)) 
to automatically choose seasonal factors. In particular, we apply fast fourier 
transform and \`{a}-trous wavelet transform~\cite{atrous}
to identify significant seasonal factors~(\S\ref{sec:tsa}). The results are used as the parameters for anomaly detection in the next step.\\
\textbf{Step $4$:} Given the time series and its seasonalities, we apply time series based anomaly 
detection~(Fig.~\ref{fig:sys}(e); Definition~\ref{def:ad}) on heavy hitter nodes. 
In particular, we apply the Holt-Winters' seasonal forecasting model~\cite{holtwinter:lisa00} to detect anomalies~(\S\ref{sec:tsa}).\\
\textbf{Step $5$:} We report anomalous events to a database. A Web-based front-end system (Fig.~\ref{fig:sys}(f)) is 
developed in JavaScript to allow SQL queries from a text database.\\
\textbf{Step $6$:} As we consider an online problem where the data arrives in a certain time period, 
we keep checking if new incoming data is arrived.

\ificdcs
\else

\subsection{Limitations}
\label{sec:problem:limitations}

Naturally, the power of \sys is limited by the operational data we have. 
We use a customer care call dataset to illustrate some of the fundamental limitations.

First, \sys can only detect network incidents that directly or indirectly affect customers. 
For example, \sys cannot detect upticks in memory utilization on aggregate 
routers if it has no sensible impact to customers. This will require a more powerful detector 
which has access to runtime memory usage.

Second, it is difficult to verify whether the customer call cases have been miscategorized. 
For example, \sys cannot locate an anomaly correctly 
if a majority of customers provide incorrect information about their geographical location.
To minimize the human error, well-trained service representatives will classify the problem based on the \emph{predefined} categories.
We believe that the selection is much less prone to error than mining natural language.

Third, \sys can only process input data which is drawn from an \emph{additive hierarchy}.
We limit our focus on hierarchical domains as it reflects the nature of the operational datasets we had in practice~(\S\ref{sec:measurement:dataset}).
The more general data representation (e.g., directed acyclic graph) is out scope of this paper and is left for future work.

Finally, \sys cannot detect anomalies where the
data rate abruptly drops, for example, link outages in network traffic data~\cite{astute:sigcomm10}. We did not consider this particular type of anomalies 
because detecting spikes (i.e., unexpectedly increases) is much more interesting in operational data like customer care calls.

\fi

%% file: algorithm.tex
\section{Online HHH Detection and Time Series Construction}\label{sec:alg}

While we presented our system overview in the previous section, it is still unclear 
how to efficiently detect the heavy hitters and construct time series in an online fashion (Step $2$ in Fig.~\ref{fig:sys}).
To motivate our design, we start by presenting a strawman proposal \bfb (\S\ref{sec:alg:baseline}).
Since the heavy hitter set may vary over timeunits, \bfb reconstructs the time series for new heavy hitters by 
traversing each timeunit within the sliding window of history.
This requires to maintain $\ell$ trees at any time. Moreover, whenever the sliding window shifts forward, 
we need to traverse all of these $\ell$ trees to construct a time series for each heavy hitter.
This is inefficient in time and space as the length of the time window $\ell$ (in terms of the timeunit size $\Delta$) is usually a large number,
resulting in an inefficient algorithm.
While this overhead can be reduced by increasing $\Delta$, or reducing the amount of history,
doing so also reduces the accuracy of inference. The typical value of $\ell$ is $8$,$064$ (a $12$-week time window that is 
long enough to accurately construct forecasting model, and a $15$-minute timeunits that is short enough to react to operational issues).
To solve this problem, we propose \alg, a novel algorithm which works by adaptively update the previous time series' position in the hierarchy
(\S\ref{sec:alg:core}). 

\subsection{A Strawman Proposal {\large \bfb}}\label{sec:alg:baseline}

\begin{algorithm}
\vspace{-4pt}
\footnotesize
    \SetAlgoLined
    \For{time instance $t = t_1, t_1+\Delta, t_1+2\Delta, \dots$}
    {
        \lIf{$(t == t_1)$}
        {
            $\kappa = \ell$;
        }
        \lElse
        {
            $\kappa = 1$;
        }\\
        \ForEach{timeunit $1\le i \le \kappa$}
        {
            Read any data item within time period $[t-i\Delta, t-i\Delta+\Delta)$ from the input stream to construct 
a tree $\textsf{Tree}[t-i\Delta]$ such that each node in the tree is associated with a count of appearances;
Notice that we use $\textsf{Tree}[\tau]$ to represent a tree constructed by the data within time period $(\tau, \tau+\Delta]$.
        }
        Do a bottom-up traversal on $\textsf{Tree}[t-\Delta]$ to derive the $\textsf{SHHH}[\theta, t-\Delta]$;\\
        \ForEach{timeunit $1\le i \le \ell$}
        {
            Given the fixed heavy hitter set $\textsf{SHHH}[\theta, t-\Delta]$, do a bottom-up traversal on $\textsf{Tree}[t-i\Delta]$ to compute time series $T[n, i]$ for each node $n \in \textsf{SHHH}[\theta, t-\Delta]$;
        }
   }
       \vspace{-10pt}
        \caption{Strawman Algorithm \bfb.}
       \vspace{-9pt}
\label{alg:strawman2}
\end{algorithm}

We first describe a strawman algorithm \bfb (Fig.~\ref{alg:strawman2}), which reuses the same data structure in different time instances, i.e., 
we destroy a tree iff it will not be used in the next time instance.
While in the first iteration (i.e., $t = t_1$) we construct $\ell$ trees, 
the time spent on the first time instance can be amortized to other time instances where we construct only one tree.
However, the time complexity of \bfb is dominated by time series construction (lines \textbf{7}-\textbf{9}).
In particular, for each time instance we need to traverse $\ell$ trees to construct a time series for each heavy hitter.

\subsection{A Low-Complexity Adaptive Scheme} \label{sec:alg:core}

In this section, we propose \alg, an adaptive scheme that greatly reduces the running time while simultaneously retaining high accuracy.
First, we present the data structure and the pseudocode (\S\ref{sec:alg:core:code}) for \alg.
In particular, \alg maintains only \emph{one} tree (rather than $\ell$ trees in \bfb), and 
only those heavy hitter nodes in the tree associated with a time series.
However, the key challenge is that heavy hitters changes over time -- 
we need efficient operations (instead of traversing $\ell$ trees) to derive the time series for heavy hitters nodes in the current time instance.
To this end, we construct time series by \emph{adapting} the time series 
generated in the previous time instance (\S\ref{sec:alg:core:sep}).
For example, consider a tree with only three nodes: a root node and its two children. 
Suppose the root node is a heavy hitter only in the current time instance, while the two children are heavy hitters only in the last time instance.
In this case, we can derive the root node's time series by merging the time series from its two children.
Although the adaptation process is very intuitive in the above example, the problem becomes more complicated as tree height increases.
We will show the proposed algorithm always finds the correct heavy hitter set (\S\ref{sec:alg:core:properties}).
While the positions of heavy hitters are perfectly correct, 
the corresponding time series might be biased after performing the \emph{split} function.
To address this problem, we will present two heuristics (\S\ref{sec:alg:core:heu1}--\S\ref{sec:alg:core:heu2}) to 
improve the accuracy. 

Without loss of generality we assume that the time instance will keep increasing by a 
constant $\varsigma \le \Delta$, where $\Delta$ is the timeunit size. 
We can always map the problem with time increment $\varsigma > \Delta$ to another equivalent problem with 
using a smaller time increment $\varsigma' \le \Delta$ such that $\varsigma'$ is a divisor 
of $\varsigma$. For simplicity, we describe our algorithm by assuming $\varsigma = \Delta$. 
We further propose a recursive procedure to consider any $\varsigma \le \Delta$ where $\varsigma$ is a 
divisor of $\Delta$
\ificdcs
\else
(\S\ref{sec:alg:core:up}).
\fi

\subsubsection{Data structure and Pseudocode} \label{sec:alg:core:code}
We maintain a tree-based data structure as follows.
Each node $n$ in the tree is associated with a type field $n.\textsf{type}$.
Fields $n.\textsf{child}[i]$ and $n.\textsf{parent}$ point to the $i$-th child and parent of the node,
$n.\textsf{depth}$ represents the depth of the node,
and $n.\textsf{weight}$ maintains the modified weight of the node (after 
discounting the weight from other descendants that are themselves heavy hitters; Definition~\ref{def:shhh}).
Array $n.\textsf{actual}$ stores the time series of the modified weights for the node $n$, and 
Array $n.\textsf{forecast}$ stores the time series of the forecasted values for the node $n$.


\begin{algorithm}
\vspace{-8pt}
\footnotesize
    \SetAlgoLined
    \For{time instance $t = t_1, t_1+\Delta, t_1+2\Delta, \dots$}
    {
        \If{$(t == t_1)$}
        {
                        Perform the same operations as lines \textbf{2}-\textbf{9} in \bfb. 
                        This gives us a tree $\textsf{Tree}$ and a heavy hitter set $\textsf{SHHH}$; 
								We set $n.$\textsf{ishh} by $true$ if $n \in \textsf{SHHH}$. Otherwise, we set $n.\textsf{ishh}$ by $false$.
								Each heavy hitter node $n \in \textsf{SHHH}$ also has a time series of actual values $n.\textsf{actual}$ and 
						      a time series of forecast values $n.\textsf{forecast}$; \\
                        \textbf{continue};\\
        }
                \ForEach{node $n \in \textsf{Tree}$}
                {
                        $n.$\textsf{washh} $\leftarrow$ $n.$\textsf{ishh}; $n.$\textsf{weight} $\leftarrow$ $0$; $n.$\textsf{tosplit} $\leftarrow$ $0$;\\
                }
                \ForEach{streaming item $(key, time)$ such that $t- \Delta \le time < t$}
                {
                        Find the node $n$ such that $n.$\textsf{type} $== key$. Increase $n.$\textsf{weight} by $1$;
                }
                Update-Ishh-and-Weight($n_{\textrm{root}}$); // Update $n.\textsf{ishh}$ and $n.\textsf{weight}$\\

                \ForEach{node $n \in \textsf{Tree}$ (bottom-up level order traversal)}
                {
                        \If{$( n.\textsf{ishh} \vee n.\textsf{tosplit} \wedge n \notin \textsf{SHHH} )$}
                        {
                                $n.\textsf{parent}.\textsf{tosplit} \leftarrow true$;\\
                        }
                }
                \ForEach{node $n \in \textsf{Tree}$ (top-down level order traversal)}
                {
                        \lIf{$( ( n \in \textsf{SHHH} \vee n.\textsf{depth}==1 ) \wedge n.\textsf{tosplit})$} 
                        {
                                SPLIT($n$);\\
                        }
                }
                \ForEach{node $n \in \textsf{Tree}$ (bottom-up level order traversal)}
                {
                        \lIf{$(n \in \textsf{SHHH} \wedge \neg n.\textsf{ishh} )$}
                        {
                                MERGE($n$);\\
                        }
                }
                \lIf{ $( n_{\textrm{root}}.\textsf{weight} < \theta )$ }
                {
                        $ \textsf{SHHH} \leftarrow \textsf{SHHH} \setminus n_{\textrm{root}} $;\\
                }
                \lElse
                {
                        $ \textsf{SHHH} \leftarrow \textsf{SHHH} \cup n_{\textrm{root}} $;\\
                }
                \ForEach{node $n \in \textsf{SHHH}$}
                {
                        Append $n.\textsf{weight}$ to $n.\textsf{actual}$; Remove the first element from $n.\textsf{actual}$ and that from $n.\textsf{forecast}$;\\
                        Derive the newest forecast value and append it to $n.\textsf{forecast}$;\\
                }
    }
       \vspace{-8pt}
    \caption{Core Algorithm of \alg. This algorithm calls three functions: Update-Ishh-and-Weight (Fig.~\ref{alg:update:ishh}), SPLIT (Fig.~\ref{alg:separation}) and MERGE (Fig.~\ref{alg:merge}).}
       \vspace{-8pt}
\label{alg:core}
\end{algorithm}

\LinesNotNumbered
\begin{algorithm}
\vspace{-8pt}
\footnotesize
\hspace{-8pt}int Update-Ishh-and-Weight(node $n$)\\
\nl\ShowLn\ForEach{node $n_c=n.\textsf{child}$}
{
        \nl\ShowLn $n.\textsf{weight} \leftarrow n.\textsf{weight} + \textrm{Update-Ishh-and-Weight}(n_c)$;\\
}
\nl\ShowLn\If{ $( n.\textsf{weight} \ge \theta)$ }
{
        \nl\ShowLn$n.\textsf{ishh} \leftarrow 1$; return $0$;\\
}
\nl\ShowLn\Else
{
        \nl\ShowLn$n.\textsf{ishh} \leftarrow 0$; return $n.\textsf{weight}$;\\
}
\vspace{-8pt}
\caption{Updating fields ${\normalfont n.\textsf{ishh}}$ and ${\normalfont n.\textsf{weight}}$. This function recursively updates ${\normalfont n.\textsf{ishh}}$ and ${\normalfont n.\textsf{weight}}$ in a bottom-up fashion.}
\vspace{-8pt}
\label{alg:update:ishh}
\end{algorithm}
\LinesNumbered

\LinesNotNumbered
\begin{algorithm}
\vspace{-8pt}
\footnotesize
        \hspace{-8pt}SPLIT(node $n$)\\
        \nl\ShowLn $\textsf{C}_n \leftarrow  \{ n_c | n_c \in n.\textsf{children}[\cdot], n_c \notin \textsf{SHHH}\}$;\\
        \nl\ShowLn \If{$( \exists n_c \in \textsf{C}_n \textrm{~s.t.~} n_c.\textsf{weight} \ge \theta   )$}
        {
                \nl\ShowLn\ForEach{node $n_c \in \textsf{C}_n$}
                {
                        \nl\ShowLn Split the time series from $n$ to $n_c$ with a scale ratio of $\textsf{F}(n_c, \textsf{C}_n )$; 
            (The function $\textsf{F}(\cdot)$ is defined in \S\ref{sec:alg:core:heu1})\\
                }
                \nl\ShowLn Clear $n.\textsf{weight}$;\\
                \nl\ShowLn $\textsf{SHHH} \leftarrow (\textsf{SHHH} \cup \textsf{C}_n) \setminus \{ n \}$;\\
        }
       \vspace{-8pt}
    \caption{Splitting time series to children.}
       \vspace{-8pt}
\label{alg:separation}
\end{algorithm}
\LinesNumbered

\LinesNotNumbered
\begin{algorithm}
\vspace{-8pt}
\footnotesize
        \hspace{-8pt}MERGE(node $n$)\\
        \nl\ShowLn \If{ ($n.\textsf{weight} < \theta$) }
        {
                {
                        \nl\ShowLn $n_p \leftarrow n.\textsf{parent}$; $\textsf{C}_n \leftarrow \{ n_c | n_c \in \{ n_p \} \cup n_p.\textsf{children}[\cdot], n_c \in \textsf{SHHH}, n_c.\textsf{weight} < \theta \}$;\\
                        \nl\ShowLn $n_p.\textsf{weight} \leftarrow \sum_{n_c \in \textsf{C}_n} n_c.\textsf{weight}$;\\
                        \nl\ShowLn $\textsf{SHHH} \leftarrow (\textsf{SHHH} \setminus \textsf{C}_n) \cup \{ n_p \}$;\\
                        \nl\ShowLn Merge the time series from every $n_c \in \textsf{C}_n$ to $n$;\\
                }
        }
       \vspace{-8pt}
    \caption{Merging time series to parent.}
       \vspace{-23pt}
\label{alg:merge}
\end{algorithm}
\LinesNumbered

Now we describe the core algorithm of \alg (Fig.~\ref{alg:core}).
We maintain a tree structure $\textsf{Tree}$ and a set of heavy hitter nodes, denoted by $\textsf{SHHH}$ over time instances.
For the first time instance (lines \textbf{2}-\textbf{5}), \alg performs the same operations as \bfb.
However, starting from the second time instance (lines \textbf{6}-\textbf{29}), \alg consists of the following three stages.\\
\textbf{Initialization} (lines \textbf{6}-\textbf{12}): We reset variables $n.\textsf{washh}$, $n.\textsf{ishh}$, and $n.\textsf{weight}$. 
        Subsequently, we recursively call a function Update-Ishh-and-Weight($n$) (Fig.~\ref{alg:update:ishh}) to calculate 
        the $n.\textsf{ishh}$ and $n.\textsf{weight}$ according to the Definition~\ref{def:shhh}.\\
        \textbf{SHHH and Time Series Adaptations} (lines \textbf{13}-\textbf{25}): 
        Unlike \bfb, where we explicitly update $\textsf{SHHH}$ using a bottom-up traversal, 
        in this stage we present a novel way to derive the set of SHHH. 
        We first perform an \emph{inverse level order traversal} (bottom-up) to check which node should ``split'' its 
        time series to the children ($n.\textsf{tosplit}$). 
        Then we perform a top-down level order traversal to split (\S\ref{sec:alg:core:sep}) the time series of 
        any node $n$ with $n.\textsf{tosplit}$ to its children.
        We then do a bottom-up level order traversal to check if we should ``merge'' (\S\ref{sec:alg:core:sep}) the time series 
        from some neighboring nodes to the parent.
        Throughout these split and merge operations, we keep updating $\textsf{SHHH}$ so that 
        any node $n \in \textsf{SHHH}$ iff $n$ has time series $n.\textsf{actual}$ and $n.\textsf{forecast}$ (except for the root node).
        Finally, we simply add/remove the root node to/from $\textsf{SHHH}$ based on its weight. 
        We will show that $\textsf{SHHH}$ satisfies the 
        Definition~\ref{def:shhh} after performing this stage (\S\ref{sec:alg:core:properties}). \\
        \textbf{Time Series Update} (lines \textbf{26}-\textbf{29}). We update the time series for each heavy hitter node in $\textsf{SHHH}$.
		  Unlike \bfb, the time series can be updated in constant time.

\subsubsection{S\hspace{1.15pt}p\hspace{1.15pt}l\hspace{1.15pt}i\hspace{1.15pt}t and M\hspace{-1.15pt}e\hspace{-1.15pt}r\hspace{-1.15pt}g\hspace{-1.15pt}e} \label{sec:alg:core:sep}

%

This section presents two functions, split and merge, to \emph{adapt} the position of heavy hitter nodes. 
Before the adaptation, heavy hitter nodes are placed in the location given in the previous timeunit. 
The goal is to move these heavy hitter nodes (with the binding time series and forecast) to 
their new positions according to the data in the current timeunit, to satisfy the Definition~\ref{def:shhh}. 

We leverage a \emph{SPLIT($n$)} function (Fig.~\ref{alg:separation}) to linearly decompose 
the time series from node $n$ into its non-heavy-hitter children. 
In other words, a heavy hitter node $n$ will be split into one or more heavy hitters in the lower level.
In this function, we first check if there exists any child node $n_c \notin \textsf{SHHH}$ with weight $\ge \theta$.
If so, we split the time series of $n$ with a scale ratio of $\textsf{F}(n_c, \textsf{C}_n)$ for each non-heavy-hitter child $n_c$
to approximate the time series of its children. We will discuss different heuristics to estimate scale ratio $\textsf{F}(n_c, \textsf{C}_n)$ (\S\ref{sec:alg:core:heu1}).


We present a \emph{MERGE($n$)} function (Fig.~\ref{alg:merge}) to combine the time series from 
one or more lower level nodes to a higher level node.
In this function, we first check if the weight of the node $n$ is smaller than $\theta$. 
If so, we merge the time series of $n$ and its siblings and parent who meet the same requirement as $n$.



\subsubsection{Correctness of Heavy Hitter Set}\label{sec:alg:core:properties} 
It suffices to show that \sys always selects the correct heavy hitter set after performing the stage ``$\textsf{SHHH}$ and Time 
Series Adaptation'' (lines \textbf{13}-\textbf{21}), i.e., 
Definition~\ref{def:shhh} will hold. The high-level intuition is that top-down split operations ensure
that there is no heavy hitter hidden in the lower levels, while bottom-up merge operations 
ensure all non-heavy-hitter nodes propagate the weight to its heavy hitter ancestor correctly.
\ificdcs
\else
The proof is given by Lemma~\ref{lem:shhh:position:ap}.

\begin{lem}\label{lem:shhh:position:ap}
{\normalfont For each time instance, after the adaptations we have $n.\textsf{ishh}$ iff $n \in \textsf{SHHH}$ for any node $n \in \textsf{Tree}$.}
\proof{
{\normalfont
We prove by induction on time instance. It is clear the argument will hold for the first time instance
as there is no adaptation. Now we show the proof will hold for any time instance.
For the ``if'' direction, suppose (by contradiction) there exists a node $n \in \textsf{SHHH}$
such that $n.\textsf{ishh}$ is $false$. Then we have $n.\textsf{weight} < \theta$ because
the function Update-Ishh-and-Weight() (Fig.~\ref{alg:update:ishh}) enforces $n.\textsf{ishh}$ iff $n.\textsf{weight} \ge \theta$.
Then by the merge rules (line \textbf{22} in Fig.~\ref{alg:core} and line \textbf{1} in Fig.~\ref{alg:merge}),
node $n$ will be merged to its parent so we have $n \notin \textsf{SHHH}$.
The only exception is the root node $n_{\textrm{root}}$ which has no parent to merge.
However, we check this special case (lines \textbf{24}-\textbf{25} in Fig.~\ref{alg:core}) to
enforce $n_{\textrm{root}}.\textsf{ishh}$ iff $n_{\textrm{root}} \in \textsf{SHHH}$.

For the ``only if'' direction, suppose (by contradiction)
there exists a node $n$ such that $n.\textsf{ishh}$ is true
and $n \notin \textsf{SHHH}$. Similarly, we have $n.\textsf{weight} \ge \theta$.
Because $n.\textsf{ishh}$ is true, $n.\textsf{parent}.\textsf{tosplit}$ is also
true (by lines \textbf{13}-\textbf{17} in Fig.~\ref{alg:core}).
We consider the following two cases. \textbf{(i)} $n.\textsf{parent}.\textsf{washh}$ is true.
In this case, by the induction hypothesis we know
$n.\textsf{parent}$ was in $\textsf{SHHH}$ before any adaptation in this time instance,
and therefore $n.\textsf{parent}$ will split its time series to $n$ (line \textbf{15} in Fig.~\ref{alg:core} and
line \textbf{2} in Fig.~\ref{alg:separation}).
If $n$ has no child, then we have $n \in \textsf{SHHH}$.
Otherwise, if $n$ splits its time series, there must exists a child $n_c$ such that $n_c.\textsf{weight} < \theta$.
If there does not exist such a child $n_c$, then we have $n.\textsf{weight}=0$ which contradicts
our previous inference $n.\textsf{weight} \ge \theta > 0$.
Because we have $n_c.\textsf{weight} < \theta$, by the merge rules (line \textbf{22} in Fig.~\ref{alg:core} and line \textbf{1} in Fig.~\ref{alg:merge})
$n_c$ will merge back the time series to $n$.
Therefore, we have $n \in \textsf{SHHH}$ after the adaptation stage.
If the time series of $n_c$ is further split into some children, similarly, there much exists a child $n_{c'}$ of $n_c$ such
that $n_{c'}.\textsf{weight} \le n_{c}.\textsf{weight} < \theta$. Then $n_{c'}$ will merge its time series to $n_c$, and then $n_c$ will merge back to $n$.
It is not hard to see we can apply this argument multiple times to make sure that $n \in \textsf{SHHH}$ after the adaptation stage.
Consider another case \textbf{(ii)} $n.\textsf{parent}.\textsf{washh}$ is false. Then we have $n.\textsf{parent} \notin \textsf{SHHH}$ before
the adaptation stage. Because $n \notin \textsf{SHHH}$, we have $n.\textsf{parent}.\textsf{weight} \ge n.\textsf{weight} \ge \theta$.
If $n.\textsf{parent}.\textsf{parent}.\textsf{washh}$ is true, then $n.\textsf{parent}.\textsf{parent}$ will split
its time series to $n.\textsf{parent}$, and then $n.\textsf{parent}$ will split its
time series to $n$ such that $n \in \textsf{SHHH}$ after the adaptation. Otherwise, if $n.\textsf{parent}.\textsf{parent}.\textsf{washh}$ is false,
we can repeatedly apply the above argument, and finally we have $n_{\textrm{root}}.\textsf{weight} \ge \theta$ and $\neg n_{\textrm{root}}.\textsf{washh}$.
Therefore, $n_{\textrm{root}}$ was not in $\textsf{SHHH}$ before the any adaptation in this time instance, and this violates the our
enforcement on root node (lines \textbf{24}-\textbf{25} in Fig.~\ref{alg:core}) in the previous time instance.
This implies the induction hypothesis is false, and we reach the contradiction.~$\square$
}
}
\end{lem}
\fi

\subsubsection{Split Rules and Error Estimation} \label{sec:alg:core:heu1}
Recall that in SPLIT($n$) function (Fig.~\ref{alg:separation}), we split the time series 
from node $n$ to each non-heavy-hitter children $n_c \in \textsf{C}_n$ with a scale ratio of $\textsf{F}(n_c, \textsf{C}_n)$, where 
$\textsf{C}_n = \{ n_c | n_c \in n.\textsf{children}[\cdot], n_c \notin \textsf{SHHH}\}$. 
The scale ratio is derived by
$\textsf{F}(n_c, \textsf{C}_n) = X_{n_c}/\sum_{\forall m \in \textsf{C}_n}X_{m}$,
where $X_n$ is a weight-related property of node $n$. Below we consider four split rules for deriving the scale ratio. 

\texttt{Uniform}: each element in the time series of $n$ is uniformly split to all its children $n_c \notin \textsf{SHHH}$, i.e., 
$X_n = 1$.

\texttt{Last-Time-Unit}: we assign $X_n$ by the weight of node $n$ in the previous timeunit. 
This assumes the weight is similar in the current timeunit and the previous one.

\texttt{Long-Term-History}: we assign $X_n$ by the total weight seen by the node across all previous timeunits.

\texttt{EWMA}: The \texttt{Last-Time-Unit} only captures the distribution in the last timeunit, while the
\texttt{Long-Term-History} treats past observations equally. This heuristic takes a different approach -- the $X_n$ is assigned by 
an exponentially smoothed weight.
 
\ificdcs
\else
We next discuss the estimation error of the ``split'' operation over time. 
For simplicity, we illustrate this by considering a EWMA-based forecast 
model $F[t] = \alpha T[i-1] + (1-\alpha) F[t-1]$, where $T[t]$ is the actual time series.
Suppose we did a split for any time series $T$ at time $t$. Also, suppose the 
forecast $F[t]$ is biased by $\xi$, i.e., the estimated forecast $F_{\mathbf{E}}[t] = F[t] + \xi$.
Then after $k$ iterations, the estimated forecast can be derived by
\begin{eqnarray}
F_{\textbf{E}}[t+k]&=& \alpha \left\{ \sum_{j=1}^{k} (1-\alpha)^{j-1} T[t+k-j] \right\} \nonumber\\
                   & & + (1-\alpha)^{k-1} (F[t] + \xi)
\end{eqnarray}
for any $k>0$. Then we have relative error 
\begin{eqnarray}
RE[t+k] &=& \frac{|F_{\textbf{E}}[t+k] - F[t+k]|}{F[t+k]}
\end{eqnarray}
As the proposed algorithm derives the forecast value in a recursive way, the inaccuracy will remain in the future forecast. 
However, the forecast error will exponentially decrease over time because of the exponential smoothing, and this is evident in Figure~\ref{fig:re}. 

\begin{figure}[t]
\centering
\includegraphics[width=3.0in]{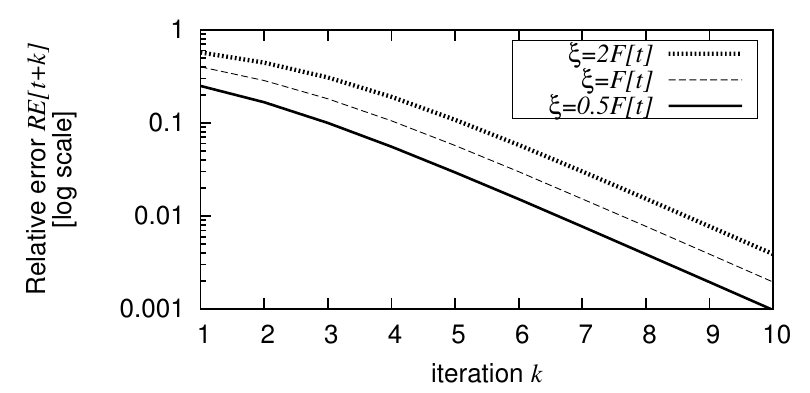}
\vspace{-8pt}
\caption{Relative error $RE[t+k]$ after $k$ iterations. (y-axis is $\log_{10}$ scale; $\alpha = 0.5$; $T[i]=1$)}
\label{fig:re}
\vspace{-10pt}
\end{figure}
\fi

\subsubsection{Reference Time Series} \label{sec:alg:core:heu2}
To further reduce the time series errors induced by the split operations, 
we propose a simple add-on to \alg by adding a set of \emph{reference time series}.
Consider a node $n$ with category $k$ in the hierarchical domain, the reference time series are 
time series of its \emph{unmodified} weight $A_n[k,t]$ (instead of $W_n[k,t]$ which 
discounts weights from heavy hitter descendants) and the time series of its forecast values.
Now we illustrate how to use the reference time series to reduce the estimation error induced by the split operations.
Suppose a node $n$ splits the time series into its children 
using a split rule. 
As split operations can be inaccurate, we assume that a child node $n_c$ received 
a ``biased'' time series $T[n_c, i]$, $1\le i \le\ell$.
If $n_c$ has a reference time series $T_{\textrm{REF}}[n_c, i]$, we will replace the $T[n_c, i]$ by 
$T_{\textrm{REF}}[n_c, i] - \sum_{n^{\star} \in \textsf{N}^{\star}} T[n^{\star}, i]$, $1\le i \le\ell$, 
where $\textsf{N}^{\star}$ is the set of $n_c$'s heavy hitter descendants, i.e., $\textsf{N}^{\star} = \{n^{\star} | n^{\star} \in \textsf{SHHH}~\wedge~n^{\star} \in desc(n_c)\}$.
We maintain the reference time series only for nodes in the top $h$ levels in the hierarchy, and 
the memory and detection accuracy tradeoffs are discussed (\S\ref{sec:eval:alg})

\ificdcs
\else
\subsubsection{Multiple Time Scales} \label{sec:alg:core:up}
We now show that \alg can support any case where $\Delta$ is a multiple of $\varsigma$ by generalizing \alg to support 
multiple time scales: consider a vector of $\eta$ time scales where the $i$-th time scale is $\lambda^{i-1} \Delta$, $1 \le i \le \eta$ for 
any two given positive integers $\lambda$ and $\eta$.

To this end, we need to maintain time series in multiple time scales. 
We use two-dimension dynamic arrays (i.e., vector$<$ vector$<$vol\_type$>$ $>$) for 
fields $n.\textsf{actual}$ and $n.\textsf{forecast}$, where 
the first dimension represents time scales.
Recall that we append the new value to the tail of time series 
in the single time scale case (line \textbf{27} in Fig.~\ref{alg:core}). 
To consider multiple timescales, we defined a recursive function UPDATE\_TS($n$,$w$,$i$) to append
a new value $w$ to the time series of node $n$ 
for the $i$-th time scale (Fig.~\ref{alg:ewmaupdate}).
In particular, when the size of any time series in the $i$-th time scale is a multiple of $\lambda$, we sum up 
the last $\lambda$ elements and recursively update the weight to the ($i+1$)-th time scale.
It is simple to see that UPDATE\_TS($n$,$w$,$i$) will be called for every $\lambda^{i}$ timeunits, for $1 \le i \le \eta$.
Then for $\kappa$ timeunits, we will call UPDATE\_TS($\cdot$) $\sum_{i=1}^{\eta} \kappa/ \lambda^{i} \le 2 \kappa$ times.
Therefore, the update operation remains $\Theta(1)$ amortized time for each time instance.

With multiple time scales, it suffices to show that \alg can support any case where $\Delta$ is a multiple of $\varsigma$.
Consider a problem $P_{A}$ with parameters $\Delta_{A}$ and $\varsigma_{A}$.
It is not hard to see that $P_{A}$ is equivalent to another problem $P_{B}$ 
with parameters $\Delta_{B} = \varsigma_{B} = \varsigma_{A}$ by assigning $\lambda_{B} = \Delta_{A} / \varsigma_{A}$ and $\eta_{B} = 1$.

\LinesNotNumbered
\begin{algorithm}
\footnotesize
        \hspace{-8pt}UPDATE\_TS(node $n$, int $w$, int $i$)\\
        \nl\ShowLn $n.\textsf{forecast}[i].\textrm{push\_back}(\alpha \times w + (1-\alpha) \times n.\textsf{forecast}[i].\textrm{last()})$;\\
        \nl\ShowLn $n.\textsf{actual}[i].\textrm{push\_back}(w)$;\\
        \nl\ShowLn $s \leftarrow n.\textsf{actual}[i].\textrm{size}()$;\\
        \nl\ShowLn \If{($i < \eta~\wedge~s = 0$ (mod $\lambda$))}
        {
                \nl\ShowLn $w' \leftarrow \sum_{1\le j \le \lambda} n.\textsf{actual}[i][s-j]$;\\
                \nl\ShowLn UPDATE\_TS($n$, $w'$, $i+1$);\\
                \nl\ShowLn \If{($s = \ell + \lambda$)}
                {
                \nl\ShowLn Do $n.\textsf{actual}[i].\textrm{pop\_head}()$ $\lambda$ times.
                }
        }
    \caption{Update time series in multiple time scales.}
\label{alg:ewmaupdate}
\end{algorithm}
\LinesNumbered
\fi

%% file: timeseries.tex
\begin{figure*}[t]
  \centering
  \subfigure[]{ \label{fig:fft:uverse}
       {\includegraphics[width=3in]{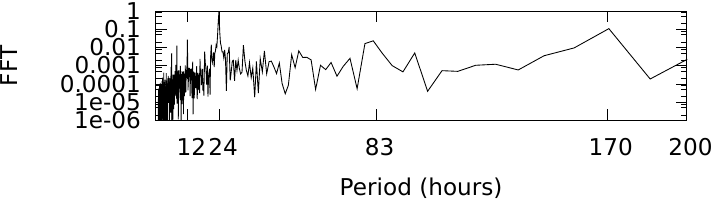}}}
  \hspace{18pt}
  \subfigure[]{ \label{fig:fft:stb}
       {\includegraphics[width=3in]{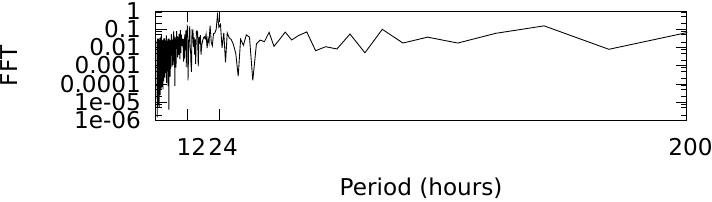}}}
  \vspace{-10pt}
  \caption{The Fast Fourier Transform for the datasets (a) \tracea and (b) \tracec. The log-scale y axis is normalized by the maximal magnitude.}
  \label{fig:fft}
\end{figure*}

\section{Time Series Analysis} \label{sec:tsa}

To study the seasonality, we apply Fast Fourier Transform (FFT) to obtain the
count of appearances in frequency spectrum (Fig.~\ref{fig:fft}). We observe that the most significant period
is $24$ hours (i.e., day period) for both datasets, while $170$ hours (i.e., the most measurable time to the weekly period) is
also noticeable in \tracea. To validate this, we also perform the \emph{\`{a}-trous wavelet transform}~\cite{atrous} to
convert the time series of our datasets to ``smoothed'' approximations at different timescales.
To avoid phase shifting, we use the low-pass $B_3$ spline
filter $(\frac{1}{16}, \frac{1}{4}, \frac{3}{8}, \frac{1}{4}, \frac{1}{16})$
as in previous work done on time series smoothing~\cite{konstantina:infocom03}.
Given the approximated time series $c_j(t)$ at timescale $j$,
the \emph{detail} of $c(t)$ at timescale $t$ can be derived by $d_j(t) = c_{j-1}(t) - c_j(t)$.
Then we have the \emph{energy} of the detail signal $d_j(t) = \sum_{\forall t}d_j(t)$,
which indicates the strength of the fluctuations at timescale $j$.
For both datasets, we apply this to multiple timescales (ranging from $5$ seconds to $4$ weeks), and
the significant periodicities we found are consistent with that given by FFT.
The results clearly suggest using time series model with \emph{multiple seasonalities} for \tracea.
We found that the periodicities of operational datasets we had are fairly stable across time.
Therefore, in our current implementation we perform the data seasonality analysis
in an offline fashion, i.e., only in the first time instance.

To distinguish anomalous behavior, we characterize typical
behavior by applying the forecasting model on time series.
As we observed strong periodicities in the datasets, simple forecasting models like exponential weighted
moving average (EWMA)~\cite{ccholt:04} will be very inaccurate.
\ificdcs
\else
To address this, we adopt the additive Holt-Winters' seasonal
forecasting model~\cite{holtwinter:lisa00}, which decouples a time series $T[t]$
into three factors: level $L[t]$, trend $B[t]$ and seasonal $S[t]$:
\begin{eqnarray}
L[t] &=& \alpha (T[t] - S[t-\upsilon]) + (1 - \alpha) ( L[t-1] + B[t-1] ) \nonumber\\
B[t] &=& \beta  (L[t] - L[t-1]       ) + (1 - \beta ) ( B[t-1]          ) \nonumber\\
S[t] &=& \gamma (T[t] - L[t]         ) + (1 - \gamma) ( S[t-\upsilon]   ) \nonumber
\end{eqnarray}
where the season length is denoted by $\upsilon$.
Then the forecast $G[t]$ is simply adding up
these three factors, i.e., $G[t] = L[t-1] + B[t-1] + S[t-\upsilon]$.
To initialize the seasonality, we assume there are at least
two seasonal cycles $T[t-1], T[t-2], \dots, T[t-2\upsilon]$.
The starting values $L[t-1], B[t-1]$ and $S[t-1], \dots, S[t-2\upsilon]$ can be derived by
\begin{eqnarray}
L[t-1] &=& \frac{1}{2\upsilon} \sum_{j=t-1}^{t-2\upsilon} T[j]  \nonumber\\
B[t-1] &=& \frac{1}{2\upsilon} \left[ \left(\sum_{j=t-1}^{\upsilon} T[j] \right) - 
\left(\sum_{j=t-\upsilon-1}^{2\upsilon} T[j] \right) \right] \nonumber\\
S[t-j] &=& T[t-j] - L[t-1], j=1,\dots,2\upsilon  \nonumber
\end{eqnarray}

The use of Holt-Winters' seasonal model does not compromise the update cost.
Given the factors in the previous timeunits, the forecast can be computed in constant time.
Moreover, we choose the additive Holt-Winters' seasonal model because of its linearity -- 
there is no need to recalculate the forecast value after merge / split.

\begin{lem}\label{lem:hwlinearity:ap}
\normalfont
(Holt-Winter Linearity).
Given $k$ Holt-Winter forecasts $G_1[t], G_2[t], G_k[t]$, where $G_i[t] = L[t-1] + B[t-1] + S[t-\upsilon]$.
Consider a time series $T^{*}[t] = \sum_{i=1}^{k}T_i[t]$ and its Holt-Winter forecast $G^{*}[t]= L^{*}[t-1] + B^{*}[t-1] + S^{*}[t-\upsilon]$.
Then $G^{*}[t] = \sum_{i=1}^{k}G_i[t]$.
\begin{proof}
We prove $L^{*}[t] = \sum_{i=1}^{k}L_i[t]$, $B^{*}[t] = \sum_{i=1}^{k}B_i[t]$ and $S^{*}[t] = \sum_{i=1}^{k}S_i[t]$ by induction on $t$.
For the initialization, we have
\begin{eqnarray}
L^{*}[t-1] &=& \frac{1}{2\upsilon} \sum_{j=t-1}^{t-2\upsilon} T^{*}[j] \nonumber\\
           &=& \frac{1}{2\upsilon} \sum_{j=t-1}^{t-2\upsilon} \sum_{i=1}^{k} T_i[j] \nonumber\\
           &=& \sum_{i=1}^{k} \left( \frac{1}{2\upsilon} \sum_{j=t-1}^{t-2\upsilon} T_i[j] \right) \nonumber\\
           &=& \sum_{i=1}^{k} L_i[t-1] \label{equ:hw:l:init}
\end{eqnarray}
\begin{eqnarray}
B^{*}[t-1] &=& \frac{1}{2\upsilon} \left[ \left(\sum_{j=t-1}^{\upsilon} T^{*}[j] \right) - 
\left(\sum_{j=t-\upsilon-1}^{2\upsilon} T^{*}[j] \right) \right] \nonumber\\
           &=& \frac{1}{2\upsilon} \left[ \left(\sum_{j=t-1}^{\upsilon} \sum_{i=1}^{k} T_i[j] \right) \right. \nonumber\\
           & & - \left. \left(\sum_{j=t-\upsilon-1}^{2\upsilon} \sum_{i=1}^{k} T_i[j] \right) \right] \nonumber\\
           &=& \sum_{i=1}^{k} \left\{ \frac{1}{2\upsilon} \left[ \left(\sum_{j=t-1}^{\upsilon} T_i[j] \right) \right. \right. \nonumber \\
           & & - \left. \left. \left(\sum_{j=t-\upsilon-1}^{2\upsilon} T_i[j] \right) \right] \right\} \nonumber\\
           &=& \sum_{i=1}^{k} B_i[t-1]
\end{eqnarray}
For $j=1, \dots, 2\upsilon$, we have
\begin{eqnarray}
S^{*}[t-j] &=& T^{*}[t-j] - L^{*}[t-1] \nonumber\\
           &=& \left( \sum_{i=1}^{k} T_i[t-j] \right) - \left( \sum_{i=1}^{k} L_i[t-1] \right) \nonumber\\ 
           &=& \sum_{i=1}^{k} \left(  T_i[t-j] - L_i[t-1] \right) \nonumber\\
           &=& \sum_{i=1}^{k} S[t-j]
\end{eqnarray}
Then we have following induction hypothesis for $j = t-1, t-2, \dots, t-2\upsilon$.
\begin{eqnarray}
L^{*}[j] &=& \sum_{i=1}^{k}L_i[j] \label{equ:hw:l:ih} \\
B^{*}[j] &=& \sum_{i=1}^{k}B_i[j] \label{equ:hw:b:ih} \\
S^{*}[j] &=& \sum_{i=1}^{k}S_i[j] \label{equ:hw:s:ih}
\end{eqnarray}
Given the induction hypothesis, now we prove the theorem would hold when $j=t$.
\begin{eqnarray}
L^{*}[t] &=& \alpha (T^{*}[t] - S^{*}[t-\upsilon]) \nonumber\\
         & & + (1 - \alpha) ( L^{*}[t-1] + B^{*}[t-1] ) \nonumber\\
         &=& \alpha \left[ \left( \sum_{i=1}^{k}T_i[t] \right) - \left( \sum_{i=1}^{k}S_i[t-\upsilon] \right) \right] + \left(1 - \alpha \right) \nonumber\\
         & & \times \left[ \left( \sum_{i=1}^{k} L_i[t-1] \right) + \left( \sum_{i=1}^{k} B_i[t-1] \right) \right]  \nonumber\\ 
         &=& \sum_{i=1}^{k} \{ \alpha \left( T_i[t] - S_i[t-\upsilon] \right) \nonumber\\
         & & + \left(1 - \alpha \right) \left( L_i[t-1] + B_i[t-1] \right) \} \nonumber\\
         &=& \sum_{i=1}^{k} L_i[t] \label{equ:hw:l:pf}
\end{eqnarray}
\begin{eqnarray}
S^{*}[t] &=& \gamma (T^{*}[t] - L^{*}[t]         ) + (1 - \gamma) ( S^{*}[t-\upsilon]   ) \nonumber\\
         &=& \gamma \left[ \left(\sum_{i=1}^{k} T_i[t] \right) - \left( \sum_{i=1}^{k} L_i[t] \right) \right] \nonumber\\
         & & + (1 - \gamma) \left( \sum_{i=1}^{k} S_i[t-\upsilon]  \right) \textrm{~(by~(\ref{equ:hw:s:ih}), (\ref{equ:hw:l:pf}))} \nonumber\\
         &=& \sum_{i=1}^{k} \left\{ \gamma (T_i[t] - L_i[t] ) + (1 - \gamma) ( S_i[t-\upsilon] ) \right\} \nonumber\\
         &=& \sum_{i=1}^{k} S_i[t]
\end{eqnarray}
\end{proof}
\label{lem:hwlinearity}
\end{lem}
\fi

%% file: evaluation.tex
\begin{table*}[t]
\caption{Summary of total running time of \sys in minutes with different algorithms. The running time for \alg is $5$-$14$ times less than that for \alg. The gap increases with the decrease of timeunit size.}
\vspace{-6pt}
\centering
\scriptsize
\begin{tabular}{c|c||c|c|c|c}\hline\hline
\multicolumn{2}{c||}{\textbf{Timeunit size ($\Delta$) in minutes} }                          & \multicolumn{2}{c|}{\textbf{$15$}} & \multicolumn{2}{|c}{\textbf{$60$}}  \\ \cline{1-6}
\multicolumn{2}{c||}{\textbf{Algorithm}}                              & \alg     & \bfb   & \alg  & \bfb          \\ \hline\hline
\multirow{2}{*}{\textsf{Reading Traces}}      & {Mean}     & $18.6~(74.0\%)  $ & $19.592~(5.4\%) $ & $19.096~(89.0\%)$  & $18.885~(16.2\%)$ \\ \cline{2-6}
                                     & {Variance} & $12.493         $ & $11.493           $ & $15.802       $  & $12.801         $ \\ \hline
\multirow{2}{*}{\textsf{Updating Hierarchies}}& {Mean}     & $4.762~(18.9\%) $ & $0.198~(0.1\%)    $ & $1.699~(7.9\%)$  & $0.273~(0.2\%) $ \\ \cline{2-6}
                                     & {Variance} & $1.082          $ & $0.082            $ & $0.287        $  & $0.116          $ \\ \hline
\multirow{2}{*}{\textsf{Creating Time Series}} & {Mean}     & $               $ & $339.216~(93.9\%) $ &                  & $96.943~(83.1\%)$ \\ \cline{2-6}
                                     & {Variance} & $               $ & $83.114           $ &                  & $86.518             $ \\ \hline
\multirow{2}{*}{\textsf{Detecting Anomalies}} & {Mean}     & $1.785~( 7.1\%) $ & $2.232~(0.6\%)    $ & $0.67~(3.1\%)$   & $0.546~(0.5\%) $ \\ \cline{2-6}
                                     & {Variance} & $0.912          $ & $2.311            $ & $0.373       $   & $0.483             $ \\ \hline\hline
\multicolumn{2}{c||}{Sum}     & $25.507~(100.0\%)$& $361.238~(100.0\%)$ & $21.465~(100.0\%)$& $116.647~(100.0\%)$\\ \hline\hline
\end{tabular}
\label{tbl:time}
\end{table*}

\section{Evaluation}
\label{sec:evaluation}

To evaluate the performance of our algorithms, we build a $\mathrm{C}$$+$$+$ implementation of \sys, including \alg and \bfb.
All our experimental results were conducted using a single core on a Solaris cluster with 2.4GHz processors.
In our evaluation, we first study \sys's system performance (\S\ref{sec:eval:alg}). Comparing it with the strawman proposal \bfb, 
the adaptive algorithm \alg provides \textbf{(i)} lower running time (by a factor of $14.2$), \textbf{(ii)} lower memory 
requirement (by a factor of $2$), \textbf{(iii)} time series with $<\negsp1\%$ absolute error, and \textbf{(iv)} identified anomalies with $99.7\%$ accuracy.
We then compare \sys with a current best practice used by an ISP operational team (\S\ref{sec:eval:dem}). 
\sys not only detects $>$$94\%$ of the anomalies found by the reference method but also many previously unknown anomalies hidden in the lower levels. 


\paragraphb{System parameters:} 
For \tracec, we use the Holt-Winters' seasonal model with single seasonal factor $S_{day}[t]$.
To capture both diurnal and weekly patterns observed in \tracea, 
we consider \emph{two} seasonal factors $S_{day}[t]$ and $S_{week}[t]$ for \tracea. 
To incorporate these two seasonal factors in the Holt-Winters' model, 
we linearly combine the factors, i.e., the seasonal factor is defined as $S = \xi S_{day}[t] + (1-\xi) S_{week}[t]$.
We choose the weight $\xi = FFT_{day} / FFT_{week} = 0.76$ according to the measured magnitude $FFT_{t}$ of frequency $1/t$.
To find the other parameters for the Holt-Winters' forecasting model, 
we measure the mean squared error of the forecast values ($\propto \sum(actual - forecast)^2$) to choose the parameters in an offline fashion.
The parameters with the minimal mean squared error are chosen for the evaluation throughout this section.
We choose a small heavy hitter threshold $\theta$, which gives us around $125$ ($5$) heavy hitters 
in the ``busy'' (``quiet'') period in \tracea, and $500$ heavy hitters in \tracec.
We perform a sensitivity test and chose sensitivity thresholds $RT=2.8$ and $DT=8$,
which gave reasonable performance in comparison with the reference method (\S\ref{sec:eval:dem}).

\subsection{System Performance} \label{sec:eval:alg}
We first evaluate the performance of \sys using algorithms \alg and \bfb. We then present the results for \tracea, and we summarize the results from \tracec.


\paragraphb{Computational Time:} 
We use \texttt{Long-Term-History} as a representative heuristic as we observed very similar running time for these heuristics.
In Table~\ref{tbl:time}, we summarize the running time of \sys using \tracea in May 2010.
We observe that \alg with $15$-minute ($1$-hour) timeunits reduces the overall running time by a factor of $14.2$ ($5.4$) as compared with \bfb.
If we subtract the running time for reading traces (which is arguably necessary for any algorithm), 
we observed a factor of $49.5$ ($41.3$) improvement.
To better understand why \alg has a better running time, we divide the program into four stages: \textsf{Reading Traces}, 
\textsf{Updating Hierarchies}, \textsf{Creating Time Series} and \textsf{Detecting Anomalies}. 
Among these stages, we observed that \textsf{Creating Time Series} is clearly a bottleneck in \bfb, 
which contributes $83.1\%$ ($93.9\%$) time of the total time of all stages for timeunits of size $15$-minute ($1$-hour). 
As the time complexity of this stage for \bfb is proportional to $\ell$, 
we expect the running time of this stage would be inversely 
proportional to the timeunit size\footnote{For a time series with fixed length, we have $\ell = 1/\Delta$.}.
Fortunately, we also observe that the ``per timeunit'' running time of this stage 
with $15$-minute timeunits is slightly smaller (with a factor of $0.87$) than that with $1$-hour time units. 
This is because the trace we used is sparse, and therefore we have a smaller tree size when using smaller timeunits. 



\paragraphb{Memory Requirement:} 
We evaluate memory costs for \alg (again, different split rules have very similar memory costs) and \bfb (Table~\ref{tbl:memory}).
To eliminate cold-start effects, we focus on measuring space costs after a long run.
We observe the memory cost of \alg is $\approx\negsp36\%$ of that of \bfb. 
We maintain the reference time series for any node in the top $h$ levels in the hierarchy (the root level is not included).
With two levels of reference nodes, the memory cost of \alg is only $\approx\negsp43\%$ of that of \bfb.

\begin{table}[!ht]
\vspace{-5pt}
\caption{Normalized memory costs of \sys with different algorithms. (Normalized memory cost $=$ total memory cost $/$ average number of nodes in the tree $/$ the memory cost of each node.)}
\vspace{-8pt}
\centering
\scriptsize
\begin{tabular}{|c|c|c|}\hline
\multirow{2}{*}{\textbf{Algorithm}} &         {\textbf{\# ref.}} &       \textbf{Normalized} \\
                                    &     {\textbf{levels ($h$)}}     &       \textbf{Space} \\ \hline\hline
\bfb                     & N/A   & $744.3$      \\ \hline
\multirow{3}{*}{\alg}    & $0$   & $271.1$  \\ \cline{2-3}
                         & $1$   & $280.2$  \\ \cline{2-3}
                         & $2$   & $322.6$  \\ \hline
\end{tabular}
\vspace{-8pt}
\label{tbl:memory}
\end{table}

\begin{figure}[t]
\centering
\vspace{-9pt}
\subfigure[]{ \label{fig:eval:error:slot}{\includegraphics[width=3.3in]{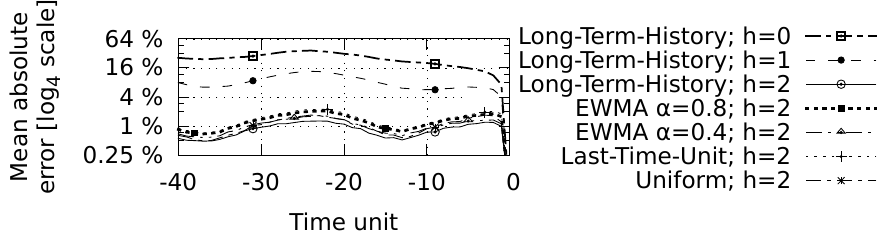} } }
\subfigure[]{\label{fig:eval:error:swindow}{\includegraphics[width=3.3in]{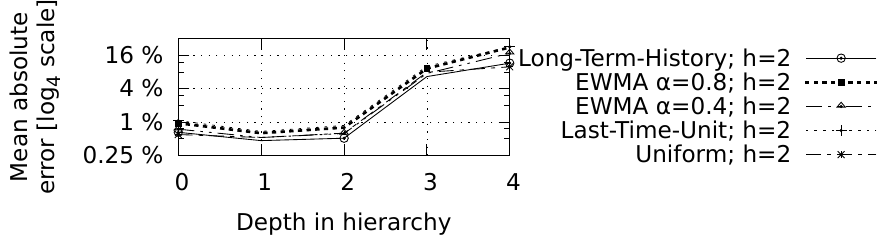}}}
\vspace{-6pt}
\caption{
The absolute error of time series for different heuristics versus (a) timeunits and (b) depths. 
The error is averaged over different timeunits in the time series and over the time series of heavy hitters. 
The reference time series are not included. The timeunit $0$ is the most recent time for any time instance.
We use $h$ and $\alpha$ to denote the number of reference levels (\S\ref{sec:alg:core:heu2}) and the smoothing rate of EWMA, respectively.
}
\vspace{-9pt}
\label{fig:eval:error}
\end{figure}

\paragraphb{Time Series Accuracy:}
As the split operations might introduce inaccuracy, we next examine how much distortion of time series we have.
We conduct the experiment by leveraging the results generated by \bfb as the ground truth. 
In particular, for each timeunit $t$ in any time series $T[t]$, 
the absolute error is derived by $|T_{\alg}[t] - T_{\bfb}[t]|$.
We first consider the impact of adding reference time series (Fig.~\ref{fig:eval:error}). 
We observe the use of the reference time series significantly reduces the error rate. 
Consider \sys with reference time series for nodes in the top $h=1$, $2$, $3$ and $4$ levels, 
we have $61$, $322$, $1$,$926$ and $45$,$479$ reference time series, respectively. 
We observe that a large number of reference time series is not necessary as two layers of reference time series ($h=2$) already gives 
us a very accurate time series estimation ($\sim$$1\%$ error in Fig.~\ref{fig:eval:error}). 
For $h=2$, we maintain reference time series for only $322/45$,$479$$\approx$$0.7\%$ of total nodes.
Moreover, we compare the accuracy for different split rules. 
We observe that only \texttt{Long-Term-History} has slightly better accuracy, 
while the accuracy results of the other heuristics are fairly close. 
Besides, we observe the accuracy is very stable over different timeunits.
While for brevity we did not present the accuracy results for every timeunit in the figure, we observe very similar results for older timeunits.

\paragraphb{Anomaly Detection Accuracy:}
We observed the heavy hitter set detected by \alg is the same as the real heavy hitter set across time.
To quantify the inaccuracy introduced by the split operation, we compare the anomalies detected by \alg and \bfb.
We use \bfb as the ground truth as it reconstructs accurate time series for each time instance.
In Table~\ref{tbl:error}, we summarize the accuracy metrics for different heuristics of \alg. 
The results are collected across $100$ time instances between May 16-21, 2010.
Overall, we found that \alg was able to detect anomalies with an accuracy of $99.7\%$ against \bfb.
However, there are some tradeoffs between split rules.
We observe that \texttt{EWMA} (with rate $\alpha=0.4$) has the highest precision, while \texttt{Uniform} has the best result on sensitivity.
We also observe \texttt{Long-Term-History} has good results on all metrics under investigation.

\paragraphb{Results for \tracec:} Compared with \tracea, it takes longer running time (with a factor of $3$) for reading traces from \tracec. 
Because of this, we observe the overall running time of \alg for \tracec is slightly increased (with a factor of $1.3$).
However, the overall running time of \bfb for \tracec is greatly increased by a factor of $7.4$. 
In particular, the running time for \textsf{Creating Time Series} is increased by a factor of $6.4$ because of the increase in the size of the hierarchy. 
Besides, we also observe an increased memory consumption for both \alg and \bfb in a 
factor of $2$ because of the larger hierarchical space. However, with $h=0$ ($h=1$) level of reference nodes, 
the memory requirement of \alg is $43\%$ ($46\%$) of that of \bfb. 
For the time series accuracy of \alg, we observed an average absolute error of only $0.8\%$ with $h=1$. 
The primary reason of high accuracy for \tracec is that 
the split operations are performed less frequently because of the smaller variance of the number of cases over time in \tracec (Fig.~\ref{fig:prop:rate:stb}).
Consequently, we observe a very good accuracy of anomaly detection for \tracec. 
In particular, we have no false positives and very few false negatives (in only about 0.13\% of all negative cases with $h=1$).

{
\begin{table}[t]
\vspace{-9pt}
\caption{Anomaly detection accuracy of Algorithm \alg. We use $h$ to denote the number of reference levels. }
\vspace{-5pt}
\centering
\scriptsize
{
\begin{tabular}{c|c||c|c|c}\hline\hline
\textbf{Split rule} & $h$ & \textbf{Accuracy} & \textbf{Precision} & \textbf{Recall} \\\hline\hline
\multirow{3}{*}{{\texttt{Long-Term-History}}}   & $0$ & $97.2\%$  & $37.8\%$  & $41.8\%$ \\ \cline{2-5}
                                     & $1$ & $98.6\%$  & $78.6\%$  & $49.3\%$ \\ \cline{2-5}
                                     & $2$ & $99.6\%$  & $94.4\%$  & $88.1\%$ \\ \hline
\texttt{EWMA} $(\textrm{rate}=0.8)$           & $2$ & $99.6\%$  & $96.5\%$  & $82.8\%$ \\ \hline
\texttt{EWMA} $(\textrm{rate}=0.6)$           & $2$ & $99.6\%$  & $96.6\%$  & $84.3\%$ \\ \hline
\texttt{EWMA} $(\textrm{rate}=0.4)$           & $2$ & $99.7\%$  & $96.7\%$  & $87.3\%$ \\ \hline
\texttt{Last-Time-Unit}                       & $2$ & $99.6\%$  & $96.5\%$  & $82.8\%$ \\ \hline
\texttt{Uniform}                              & $2$ & $99.3\%$  & $79.2\%$  & $94.0\%$  \\ \hline\hline
\end{tabular}
\vspace{-8pt}
\label{tbl:error}
}
\end{table}
}

\begin{table}[!t]
\caption{A comparison of \alg against the reference method}
\vspace{-5pt}
\centering
\scriptsize
\begin{tabular}{|c|c|}\hline
\textbf{Performance metric} & \textbf{Value} \\\hline\hline
       {Type 1 (Accuracy) $=\frac{\#\textrm{TPs}+\#\textrm{TNs}}{\#\textrm{Cases}}$ }     & $94.1\%$   \\\hline
       {Type 2 $=\frac{\#\textrm{TPs}}{\#\textrm{TPs}+\#\textrm{MAs}}$}     & $90.9\%$   \\\hline
       {Type 3 $=\frac{\#\textrm{TNs}}{\#\textrm{TNs}+\#\textrm{NAs}}$}     & $94.1\%$   \\\hline
\end{tabular}
\vspace{-12pt}
\label{tbl:cmplarry}
\end{table}

\subsection{Comparison with Current Practice} \label{sec:eval:dem}
To evaluate the performance of \sys in identifying live
network problems, we compared the set of anomalies detected by our
methods from \tracea with a set of reference anomalies identified using an existing approach 
based on applying control charts to time series of aggregates at
the first network level (the VHO level). 
This existing approach is not scaled to lower levels and is currently used by an operational team of a major US broadband provider.

For the purposes of this study we treat the reference anomalies as the ground truth
against which we evaluate the performance of our method. 
Unfortunately, the reference anomaly set is \emph{not complete} -- it only 
looks at the anomalies that happened in the first network level (VHO).
Because of this, we cannot simply use the standard terms \{True$|$False\} \{Positive$|$Negative\} in describing 
the anomalies found by our method relative to the reference set.
To undertake the comparison, we define the following comparison metrics.
Let $A_{\opg}$ and $A_{\sys}$ denote the set of reference anomalies
and those found by \sys, respectively. 
Clearly, each anomaly $a$ can be one-to-one mapping to a certain network location $L(a)$ and timeunit $T(a)$.
In particular, for each anomaly $a_{\opg} \in A_{\opg}$, we have a \emph{true alarm} (TA) case 
if there exists an anomaly $a_{\sys} \in A_{\sys}$ such that $T(a_{\opg}) = T(a_{\sys})$ and 
$L(a_{\opg}) \sqsupseteq L(a_{\sys})$, where $L_1 \sqsupseteq L_2$ iff $L_1$ is equal to or an ancestor of $L_2$.
In this case, \sys is able to locate the anomaly with finer granularity.
For each anomaly in $A_{\opg}$, we have a \emph{missed anomaly} (MA) case 
if there does not exist any anomaly satisfying the above requirement. 
Moreover, let $A^{I}_{\sys} \subset A_{\sys}$ denote the set of \sys's
anomalies which is not related to any anomaly $a_{\opg} \in A_{\opg}$.
Specifically, $A^{I}_{\sys}$ $=$ $\mathcal{G}(A_{\sys}, A_{\opg})$ $=$ $A_{\sys}$ $\setminus$ $\left\{ a_{\sys} \in A_{\sys} | \exists a_{\opg} 
\in A_{\opg}, T(a_{\opg}) = T(a_{\sys}),\right.$\\$\left. L(a_{\opg}) \sqsupseteq L(a_{\sys}) \right\}$. 
We say there is a \emph{new anomaly} (NA) case for each element in $A^{I}_{\sys}$.
Let $\hat{A}_{\sys}$ denote the set of heavy hitters which are not reported as anomalies by \sys.
For each element in $\hat{a}_{\sys} \in \hat{A}_{\sys}$, 
we have a \emph{true negative} (TN) case if $\hat{a}_{\sys}$ is unrelated to any anomaly in $A_{\opg}$. 
In particular, the set of TN cases is defined by $\mathcal{G}(\hat{A}_{\sys}, A_{\opg})$, 
where the function $\mathcal{G}(\cdot)$ is defined in the NA case.

We compare the reported anomalies using \tracea in the period of
$14-24$th, September $2010$.  In Table~\ref{tbl:cmplarry}, we observe
that \alg effectively detects most of the anomalies found by the reference method with an accuracy of $94\%$, i.e., we have a large number of 
true alarms and true negatives, while the number of missed anomaly is very small.

\ificdcs
\else
Furthermore, the results suggest that \sys is promising to find previous unknown 
anomalies. This is likely attributable to the fact that \sys is able to adapt heavy hitter positions to detect 
anomalies below the first level, whereas the reference method is only able to trace the first network level nodes.
We perform a simple data aggregation of the NAs to remove any redundant 
anomalies which are an ancestor of other anomalies. The fraction of the resulting cases 
in the (VHO, IO, CO, DSLAM) level is ($5\%$, $56.3\%$, $29.3\%$, $9.4\%$).
The majority ($95\%$) of NA cases are localized below the VHO level; 
These are inherently harder to detect with the
reference dataset due to their VHO-level aggregation with non-anomalous call loads.
\fi

%% file: related.tex
\section{Related work}
\label{sec:related}

Given the increasing severity of DoS, scanning, and other malicious traffic,
traffic anomaly detection in large networks is gaining increased attention.  To
address this, several works detect statistical variations in traffic matrices
constructed using summarizations of traffic feature distributions~\cite{Cormode:05:ja,Estan:sigcomm02}.  
Statistical methods used in these works include sketch-based approach~\cite{Krishnamurthy:imc03}, ARIMA
modeling~\cite{ARIMA:1994}, Kalman filter~\cite{Soule:imc05} and wavelet-based methods~\cite{Barford:imw01}.
ASTUTE~\cite{astute:sigcomm10} was recently proposed to achieve high detection 
accuracy for correlated anomalous flows. 
Lakhina \textit{et al.}~\cite{lakhina:sigcomm04, lakhina:sigcomm05} showed that Principal
Component Analysis (PCA) can improve scaling by reducing the dimensionality over 
which these statistical methods operate. 
Huang \textit{et al.}~\cite{Huang:infocom07} proposed a novel 
approximation scheme for PCA which greatly reduce the 
communication overheads in a distributed monitoring setting. 
More recently, Xu \textit{et al.}~\cite{Xu:sosp09} also applied 
PCA to mine abnormal feature vectors in console logs.
However, it is unclear how to transform hierarchical domain in 
operational data to multiple dimensions for PCA. 
While entropy of data distribution might be a useful measure as 
suggested in previous studies~\cite{lakhina:sigcomm05, Nychis:imc08}, 
Ringberg \textit{et al.}~\cite{Ringberg:sigmetrics07} showed that the choice of aggregation can 
significantly impact the detection accuracy of PCA.

There are several works on hierarchical heavy hitter detection~\cite{Cormode:sigmod04, autofocus, Ajay:sigcomm09, Zhang:imc04}.
Among these, our earlier work~\cite{Zhang:imc04} on hierarchical heavy hitter detection 
is the closest in spirit to \sys. 
However, \textsf{HHD} assumes \emph{cash-register} model~\cite{Muthukrishnan:soda03},
where the input stream elements cannot be deleted~\cite{Cormode:sigmod04}.
Therefore, \textsf{HHD} is more useful to detect \emph{long-term} heavy hitters over a coarser time
granularity. 
To detect heavy hitters on the most recent time period, 
our strawman proposal \bfb can be seen as a nature extension of \textsf{HHD} where 
we apply \textsf{HHD} for every timeunit.

%% file: conclusion.tex
\section{Conclusion}
\label{sec:conclusion}


We develop \sys to detect anomalies designed toward the unique characteristics of operational data,
including a novel adaptive scheme to accurately locate the anomalies 
using hierarchical heavy hitters.
Using the operational data collected at a tier-1 ISP, our evaluation suggests that \sys maintains accuracy 
and computational efficiency that scales to large networks.
We deploy our implementation on a tier-1 ISP, and our major next step is to optimize system parameters 
based on the feedback from ISP operational teams.
